\documentclass[12pt,a4paper]{article}
\usepackage[utf8]{inputenc}
\usepackage[english]{babel}
\usepackage{graphicx, color}
\usepackage{amsmath, amsfonts, amssymb, bm}

\usepackage[top=2.5cm, bottom=2cm, left=2cm, right=2cm]{geometry}

\title{High-field mobility in graphene on substrate with a proper inclusion of the Pauli exclusion principle\thanks{This work has been partially supported by the University of Catania, project F. I. R.  {\em Charge transport in graphene and low dimensional systems},  and by INDAM.}}

\author{Marco Coco\thanks{mcoco@dmi.unict.it} \and Armando Majorana\thanks{majorana@dmi.unict.it} \and Giovanni Nastasi\thanks{g.nastasi@unict.it} \and Vittorio Romano\thanks{romano@dmi.unict.it}}

\date{Department of Mathematics and Computer Science\\
University of Catania}

\newcommand{\te}{\mathrm{t}}
\newcommand{\bx}{\mathbf{x}}
\newcommand{\bk}{\mathbf{k}}
\newcommand{\eps}{\varepsilon}

\newcommand{\sv}{\, ,}
\newcommand{\p}{\, .}
\newcommand{\eq}[1]{(\ref{#1})}
\newcommand{\dm}{\displaystyle}

\begin{document}
\baselineskip=20pt
\noindent

\maketitle

\begin{abstract}
The aim of this work is to simulate the charge transport in  a monolayer graphene on different substrates. This requires the inclusion of the 
scatterings of the charge carriers with the impurities and the phonons of the substrate, besides the interaction mechanisms already present
in the graphene layer.   
 As physical model, the semiclassical Boltzmann 
equation is assumed and the results are based on Direct Simulation Monte Carlo (DSMC).  A crucial point is the correct inclusion of the Pauli Exclusion Principle (PEP). Most simulations use the approach proposed in \cite{Lugli} which, however, predicts an occupation number greater than one with an evident violation of PEP.  Here the Monte Carlo scheme devised in \cite{RoMaCo} is employed. It predicts occupation numbers consistent with PEP and therefore is physically more accurate. 

Two different substrates are investigated: SiO$_2$ and hexagonal boron nitride (h-BN). The model in \cite{Hwang2007b}  for the charge-impurities scattering has been adopted. In such a model a crucial parameter is the distance $d$ between the graphene layer and the impurities of the substrate.  Usually  $d$ is considered constant \cite{CoMajRo}. Here we assume that $d$ is a random variable in order to take into account the roughness of the substrate and the randomness of the location of the impurities. 

We confirm, as in \cite{Hirai},  where only the low-field mobility has been investigated,  that h-BN is one of the most promising substrate also for the high-field mobility on account of  the reduced degradation of the velocity due to the remote impurities.
%\keywords{Graphene \and Direct  Simulation Monte Carlo \and Charge transport}
%% \PACS{PACS code1 \and PACS code2 \and more}
%\subclass{82C70 \and 82C80 \and 65M60}
\end{abstract}

\section{Introduction}
\label{intro}
\begin{figure}[!htb]
\begin{center}
\includegraphics[width=0.5\textwidth]{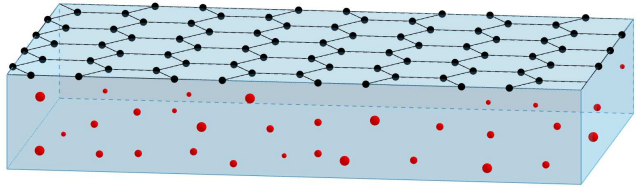} 
%\begin{spacing}{1.0}
\caption{The graphene sheet over a substrate. The spheres represents the impurities. 
% include label in command for correct ref count
\label{gr_sh}}
%\end{spacing}
\end{center}
\end{figure}

Graphene is a gapless semiconductor made of a single layer of carbon 
atoms arranged into a honeycomb hexagonal lattice. Around the Dirac points,
it has, as first approximation, a conical band structure, so electrons have a zero 
effective mass and they exhibit a photon-like behavior. 
A physically accurate model for charge transport is given by a semiclassical Boltzmann 
equation whose scattering terms have been deeply analyzed in the last decade.
Quantum effects has also been included in the literature but for  Fermi energies high enough, as those considered in this paper,  
the interband tunneling effect is practically negligible and the semiclassical approach reveals satisfactory \cite{KaLaMau}.
The aim of this work is to simulate a monolayer graphene on a substrate, as, for 
instance, considered in \cite{Hirai}  (see 
Figure \ref{gr_sh}), at variance with the case of suspended graphene studied in 
\cite{RoMaCo}.
\\
Usually the available solutions are
obtained with direct Monte Carlo simulations.  The peculiar band structure of graphene requires that the Pauli exclusion principle must be taken into account but  the standard Monte Carlo
approaches suffer from a violation of such a principle because they predict a  maximum occupation number grater than one.  In \cite{RoMaCo} a new Direct Simulation Monte Carlo (DSMC) procedure has been devised in order to overcome such a difficulty and successfully applied to charge transport in suspended monolayer graphene. Comparison with  direct solutions of the electron Boltzmann equation obtained with Discontinuous Galerkin (DG)  methods \cite{Cockburn} have confirmed the validity of the approach \cite{CoMajRo}. 

 Apart from the scatterings already present in the suspended case, now  also the effects of the remote phonons  and the impurities of the substrate must be included. The scattering rate between the electrons and the phonons of the substrate is similar to that of the suspended case while the interaction with the impurities adds noticeable additional difficulties, mainly due to the rather involved expression of the dielectric function which is itself a source of theoretical debates \cite{Hwang2007b,Hwang2007}.    

We will assume the model proposed in \cite{Hwang2007b}  for the charge-impurities scattering. A crucial parameter is the depth $d$ of the remote impurities. It is of the order of a few angstroms but the exact value can vary from a specimen to another. In \cite{CoMajRo} $d$ has considered constant and the results for several values of $d$ have been compared.  

Here we take into account the randomness of the impurities location, related also to the roughness of the interface of the oxide, by considering $d$ a random variable. Various  distributions have been analyzed: uniform and chi-square with several degrees of freedom.  

Two different substrates have been tackled: SiO$_2$ and hexagonal boron nitride (h-BN).  In  \cite{Hirai}   HfO$_2$ has been also considered but the analysis at low fields  reveals that it is not an adequate  material because the strong degradation of the mobilities. 
Our analysis confirms that the h-BN is a better material than SiO$_2$  on account of the reduced degradation of the mobility and the stability  with respect to the fluctuations of the parameter $d$, even if significant quantitative differences are found with respect to \cite{Hirai}. h-BN assures the higher mobility and its performance  is robust with respect to the randomness of $d$.

The plan of the paper is as follows. In section 2 the semiclassical kinetic model for charge transport in graphene on a substrate is outlined. In section 3 the DSMC is discussed and in the last section  the numerical results are presented and commented. 

\section{Semiclassical charge transport in graphene on a substrate}
\label{sec:2}
In a semiclassical kinetic setting, the charge transport in graphene is described by 
four Boltzmann equations, one for electrons in the valence ($\pi$) band and one for 
electrons in the conductions ($\pi^*$) band, that in turn can belong to the $K$ or $K'$ 
valley,
\begin{equation}
\frac{\partial f_{\ell,s}(\te,\bx,\bk) }{\partial \te} + {\bf v}_{\ell,s} \cdot 
\nabla_{{\bx}} 
f_{\ell,s}(\te,\bx,\bk)  - \frac{e}{\hbar} {\bf E} \cdot \nabla_{\bk} 
f_{\ell,s}(\te,\bx,\bk)
= \left( \dfrac{df_{\ell,s}}{d \te}(\te,\bx,\bk) \right)_{coll} ,
\label{transport}
\end{equation}
where $f_{\ell,s}(\te,\bx,\bk)$ represents the distribution function of charge carriers in 
the valley $\ell$  ($K$ or  $K'$), band  $\pi$ or $\pi^*$ ($s = -1$ or $s=1$) at position 
${\bx}$, time $\te$ and wave-vector ${\bk}$. 
We denote by $\nabla_{\bx}$ and $\nabla_{\bk}$ the gradients with respect to the position 
and wave-vector, respectively.
The microscopic velocity ${\bf v}_{\ell,s}$ is related to the energy band $\eps_{\ell,s}$ 
by 
$$
{\bf v}_{\ell,s} = \dfrac{1}{\hbar} \, \nabla_{\bk} \, \eps_{\ell,s} \p
$$
With a very good approximation \cite{CaNe} a linear dispersion relation holds for the 
energy bands  $\eps_{\ell,s}$ around the equivalent Dirac points; so that
$\eps_{\ell,s} = s \,  \hbar \, v_F \left| \bk -\bk_{\ell} \right| $, where $v_F$ is the 
(constant) Fermi velocity, $\hbar$ is the Planck constant divided by $2 \, \pi$, and 
$\bk_{\ell}$ is the position of the Dirac point $\ell$ in the first Brillouin zone.
The elementary (positive) charge is denoted by $e$, and ${\bf E}$ is the electric field, 
here assumed to be constant. 
The right hand side of Eq.~\eq{transport} is the collision term representing the 
interaction of electrons with impurities and phonons, the latter due to both the 
graphene crystal and substrate \cite{Fang}. 
Acoustic phonon scattering is intra-valley and intra-band. Optical phonon scattering is intra-valley 
and can be  longitudinal optical ($LO$) and transversal optical ($TO$); it can be intra-band or 
inter-band. Scattering with optical phonons of type $K$  pushes electrons from a valley to the other 
one (inter-valley scattering). 
In addition to the interactions already present in the suspended case, surface optical phonon scattering and charged impurity 
(imp) scattering 
induced by the substrate are also included. 

We assume that phonons are at thermal equilibrium.
The general form of the collision operator can be written as
\begin{eqnarray*}
\left( \dfrac{d f_{\ell,s}}{d \te}(\te,\bx,\bk) \right)_{coll}
&=& \sum_{\ell', s'} \left[
\int S_{\ell', s', \ell, s}(\bk', \bk) \,
f_{\ell',s'}(\te,\bx,\bk') \left( 1 - f_{\ell,s}(\te,\bx,\bk) \right) d \bk'  \right.
\\
&&
\left.  -
\int S_{\ell, s, \ell', s'}(\bk, \bk') \,
f_{\ell,s}(\te,\bx,\bk) \left( 1 - f_{\ell',s'}(\te,\bx,\bk') \right) d \bk' \right] 
\end{eqnarray*}
where the transition rate $\dm S_{\ell', s', \ell, s}(\bk', \bk)$ is given by the sum of 
terms of kind
\begin{eqnarray}
&&
\left| G^{(\nu)}_{\ell', s', \ell, s}(\bk', \bk) \right|^{2} \left[ 
\left( n^{(\nu)}_{\mathbf{q}} + 1 \right)
\delta \left( \eps_{\ell,s}(\bk) - \eps_{\ell',s'}(\bk') + \hbar \, \omega^{(\nu)}_{\mathbf{q}}
\right) \right. \nonumber
\\
&&
\left. \mbox{} \hspace{100pt} + n^{(\nu)}_{\mathbf{q}} \,
\delta \left( \eps_{\ell,s}(\bk) - \eps_{\ell',s'}(\bk') - \hbar \, \omega^{(\nu)}_{\mathbf{q}}
\right) \right]   \label{Scatt}
\end{eqnarray}
related to electron-phonon scatterings and other terms corresponding to the scatterings with 
the impurities.

The index $\nu$ labels the $\nu$th phonon mode, $G^{(\nu)}_{\ell', s', \ell, s}(\bk', \bk)$
is the kernel, which describes the scattering mechanism, due to phonons $\nu$, of 
electrons belonging to valley $\ell'$ and band $s'$, and electrons belonging to valley $\ell$ and 
band $s$.
The symbol $\delta$ denotes the Dirac distribution, $\omega^{(\nu)}_{\mathbf{q}}$ is the 
$\nu$th phonon frequency, $n^{(\nu)}_{\mathbf{q}}$ is the Bose-Einstein distribution for the 
phonon of type $\nu$
$$
n^{(\nu)}_{\mathbf{q}} = \dfrac{1}{e^{\hbar \, \omega^{(\nu)}_{\mathbf{q}} /k_B T} - 1} \sv
$$
$k_B$ is the Boltzmann constant and $T$ the constant graphene lattice temperature.
When, for a phonon $\nu_{*}$, $\hbar \, \omega^{(\nu_{*})}_{\mathbf{q}} \ll k_B T$, the 
scattering with the phonon $\nu_{*}$ can be assumed elastic. 
In this case, we eliminate in Eq.~\eq{Scatt} the term 
$\hbar \, \omega^{(\nu_{*})}_{\mathbf{q}}$ inside the delta distribution and we use the 
approximation 
$n^{(\nu_{*})}_{\mathbf{q}} \approx k_B T / \hbar \, \omega^{(\nu)}_{\mathbf{q}} - \frac{1}{2}$.

We will describe the terms of the collision operator concerning the scatterings with the impurities 
in the sequel.
\subsection{The model with only one distribution function}

By applying a gate voltage transversal with respect to the graphene sheet, it is possible to modify the Fermi energy $\varepsilon_F$ and therefore the charge density.
If a high positive value of the 
Fermi energy is considered, the electrons responsible for the current are those belonging to the conduction band. Therefore, only the transport equation for electrons in the 
conduction band is considered and interband electron transitions are neglected.  Moreover 
the valleys $K$ and $K'$ are considered as equivalent.
A reference frame centered in the $K$-point  will be used. 
Of course, we simplify the notation, omitting the indexes $s$ and $\ell$ and denoting by $f$ the only relevant distribution function.

The expressions of the electron-phonon scattering matrices  used in our simulations are as follows.
\\
For acoustic phonons, usually one considers the elastic approximation, and therefore
\begin{equation}
\left( 2 \, n^{(ac)}_{\mathbf{q}} + 1 \right)
\left| G^{(ac)}(\bk', \bk) \right|^{2} = \dfrac{1}{(2 \, \pi)^{2}} \, 
\dfrac{\pi \, D_{ac}^{2} \, k_{B} \, T}{2 \hbar \, \sigma_m \, v_{p}^{2}}
\left( 1 + \cos \vartheta_{\bk \sv \bk'} \right) ,
\label{transport_acoustic}
\end{equation}
where $D_{ac}$ is the acoustic phonon coupling constant, $v_{p}$ is the sound speed in graphene, 
$\sigma_m$ the graphene areal density, 
and $\vartheta_{\bk \sv \bk'}$ is the convex angle between $\bk$ and ${\bk'}$.   
\\
There are three relevant optical phonon scatterings: the longitudinal optical ($LO$), the transversal 
optical ($TO$) and the ${K}$  phonons.
The electron-phonon scattering matrices are
\begin{eqnarray}
\left| G^{(LO)}(\bk', \bk) \right|^{2} & = & 
\dfrac{1}{(2 \, \pi)^{2}} \, \dfrac{\pi \, D_{O}^{2}}{\sigma_m \, \omega_{O}}
\left( 1 - \cos ( \vartheta_{\bk \sv \bk' - \bk} + \vartheta_{\bk' \sv \bk' - \bk} ) \right)
\label{GLO}
\\
\left| G^{(TO)}(\bk', \bk) \right|^{2} & = & 
\dfrac{1}{(2 \, \pi)^{2}} \, \dfrac{\pi \, D_{O}^{2}}{ \sigma_m \, \omega_{O}}
\left( 1 + \cos ( \vartheta_{\bk \sv \bk' - \bk} + \vartheta_{\bk' \sv \bk' - \bk} ) \right)
\label{GTO}
\\
\left| G^{(K)}(\bk', \bk) \right|^{2} & = & 
\dfrac{1}{(2 \, \pi)^{2}} \, \dfrac{2 \pi \, D_{K}^{2}}{\sigma_m \, \omega_{K}}
\left( 1 - \cos \vartheta_{\bk \sv \bk'} \right) ,
\label{GK}
\end{eqnarray}
where $D_{O}$ is the optical phonon coupling constant, $\omega_{O}$ is the optical phonon frequency,
$D_{K}$ is the K-phonon coupling constant and $\omega_{K}$ is the $K$-phonon frequency.
The angles $\vartheta_{\bk \sv \bk' - \bk}$ and $\vartheta_{\bk' \sv \bk' - \bk}$ denote 
the convex angles between $\bk$ and $\bk' - \bk$  and between $\bk'$ and  $\bk' - \bk$, 
respectively. \\
Due to the presence of  the substrate, we must also include the interactions between the electrons of 
the graphene sheet and the remote phonons and impurities of the substrate.
The electron-phonon scattering matrices  have the same form of \eq{GLO} and 
\eq{GTO}. 
Regarding the remote impurity scattering, we assume that they stay in a plane at distance 
$d\, \, $ from the graphene sheet.
The definition of the transition rate for electron-impurity scattering is highly complex; so many 
approximate models are proposed.  
Following \cite{Hwang2007b}, we adopt the transition rate 
\begin{equation}
S^{(imp)}(\bk, \bk') = \dfrac{2 \pi}{\hbar}  \, \dfrac{n_{i}}{(2 \, \pi)^{2}} 
\left| \dfrac{V_{i}(|\bk - \bk'|, d)}{\epsilon(|\bk - \bk'|)} \right|^{2}
\dfrac{\left(1 + \cos \vartheta_{\bk \sv \bk'} \right)}{2}
\delta \left( \eps(\bk') - \eps(\bk) \right) ,
\end{equation}
where
\begin{itemize}
\item[a)] 
$n_{i}$  is the number of impurities per unit area. 
\vskip 0.3cm
\item[b)]
$\dm V_{i}(|\bk - \bk'|, d) = 2 \, \pi e^{2} \, 
\dfrac{\exp(- \, d \, |\bk - \bk'|)}{\tilde{\kappa} \, |\bk - \bk'|}$ 
\begin{itemize}
\vskip 0.3cm
\item 
$d$ is the location of the charged impurity measured from the graphene sheet 
\item
$\dm \tilde{\kappa}$ is the effective dielectric constant, defined by
$\dm 4 \pi  \epsilon_{0} \left( \kappa_{top} + \kappa_{bottom}  \right) / 2$, where $\epsilon_{0}$ 
is the vacuum dielectric constant and $\kappa_{top}$ and $\kappa_{bottom}$ are the relative 
dielectric constants of the medium above and below the graphene layer. 
For example, if the materials are SiO$_{2}$ and air one has
$\dm \tilde{\kappa} = 4 \pi  \epsilon_{0}  \left( 1 + \kappa_{SiO_{2}} \right) / 2 \approx 4 \pi \times 2.45 \, 
\epsilon_{0}$. For the other material see Table  \ref{Cost_fis_sub}.
\end{itemize}
\vskip 0.3cm
\item[c)]
\begin{small}
$\dm \epsilon(|\bk - \bk'|) = \left\{
\begin{array}{ll}
 1 + \dfrac{q_{s}}{|\bk - \bk'|} - \dfrac{\pi \, q_{s}}{8 \, k_{F}} \hfill
  \mbox{if  \,\,} |\bk - \bk'| < 2 \, k_{F} &
\\[15pt]
 1 + \dfrac{q_{s}}{|\bk - \bk'|} - 
 \dfrac{q_{s} \sqrt{|\bk - \bk'|^{2} - 4 \, k_{F}^{2}}}{2 \, |\bk - \bk'|^{2}} -
 \dfrac{q_{s}}{4 \, k_{F}} \, \mbox{asin} \left( \dfrac{2 \, k_{F}}{|\bk - \bk'|} \right)
 \hfill \mbox{otherwise}&
\end{array}
\right. 
$
\end{small}
\vskip 0.3cm
is the 2D finite temperature static random phase approximation (RPA) dielectric (screening) function appropriate for 
graphene;
\begin{itemize}
\item 
$\dm q_{s} = \dfrac{4 \, e^{2} \, k_{F}}{\tilde{\kappa} \, \hbar \, v_{F}}$
is the effective Thomas-Fermi wave-vector for graphene; it can be rewritten in terms of the dimensionless Wigner-Seitz radius $ r_S$ as
$\dm q_{s} = 4 r_S k_F$;
\item
$\dm k_{F} = \dfrac{\varepsilon_F}{\hbar v_F}$  is the Fermi wave-vector.
\end{itemize}
\end{itemize}
$d$ is usually fixed once for all in each simulation. However, it is more realistic to assume that $d$ can vary because the impurities are implanted with a certain degree of uncertainty in their location. 
As already shown in \cite{CoMajRo}, $d$ is crucial for a correct prediction of the electron velocity, and therefore, in turn, of the electron  mobilities. In the present paper we assume that $d$ is a random variable. A uniform distribution and chi-square distributions with several degree of freedom will be considered. The choice of the latter ones is due to their flexibility
to model the distance of the impurities with a realistic unlikely value of $d$ close to zero.

We close this section evaluating the transition rates (collision frequencies) associated to the scattering mechanisms introduced above. 
For the $A$th type of scattering the  transition rate is defined as
$$
\Gamma_A (\bk) =  \int S_A (\bk, \bk') \: d \bk' 
$$
and depends on $\bk$ only through the energy, that is indeed $\Gamma_A (\bk) = \Gamma_A  (\eps)$.

For the acoustic phonon scattering we get
\begin{eqnarray*}
\Gamma_{ac}  (\eps) = \dfrac{ D_{ac}^{2} \, k_{B} \, T}{4 \hbar^3  \,  v_F^2 \, \sigma_m \, v_{p}^{2}} \,\, \eps \label{rate_ac}
\end{eqnarray*}
while for the total optical phonon scattering, given by the sum of the longitudinal and transversal contribution,  we have
\begin{eqnarray*}
\Gamma_{op}  (\eps) = \dfrac{D_{O}^{2}}{ \sigma_m \, \omega_{O} \hbar^2 \, v_F^2} \left[\left(\eps -   \hbar \, \omega_{O} \right) \left( n^{(O)}_{\mathbf{q}} + 1 \right) H (\eps  +   \hbar \, \omega_{O}) +        \left(\eps + \hbar \, \omega_{O} \right) \, n^{(O)}_{\mathbf{q}} \right], \label{rate_op}
\end{eqnarray*}
where the fact that the coupling constants are the same for  both the longitudinal and the transversal optical phonons has been used. In Eq.~\eq{rate_op} $H$ is the Heaviside function and $n^{(O)}_{\mathbf{q}} $ the equilibrium optical phonon distribution. 

The expression of the transition rate for the $K$ phonon scattering is the same as for the optical phonon 
\begin{eqnarray*}
\Gamma_{K}  (\eps) = \dfrac{D_{K}^{2}}{ \sigma_m \, \omega_{K} \hbar^2 \, v_F^2} \left[\left(\eps -   \hbar \, \omega_{K} \right) \left( n^{(K)}_{\mathbf{q}} + 1 \right) H (\eps -   \hbar \, \omega_{K}) +        \left(\eps + \hbar \, \omega_{K} \right) \, n^{(K)}_{\mathbf{q}} \right]. \label{rate_K}
\end{eqnarray*}
Above $n^{(K)}_{\mathbf{q}} $ is the equilibrium $K$ phonon distribution.
At last the transition rate for the impurity scattering, due to the rather involved expression, has to be evaluated numerically. Following a standard procedure, the following correction is adopted 
\cite{Lund}
\begin{eqnarray}
\Gamma_{imp}  (\bk) = \int S^{(imp)}   (\bk, \bk') \left( 1 - \cos \vartheta_{\bk \sv \bk'} \right)  \: d \bk'.
\end{eqnarray}
The physical parameters for the scattering rates are summarized in Tables \ref{tabella}, \ref{Cost_fis_sub} and are the same as in \cite{Hirai}.
\begin{table}[h]
\caption{Physical parameters for the scattering rates in pristine graphene.}
\label{tabella}
\centering
\begin{tabular}{||l|c||l|c||}
\hline & & & \\[-10pt]
$v_{F}$ & $10^{8}$ cm/s & $v_{p}$ & $2 \times 10^{6}$ cm/s \\[2pt]
\hline & & & \\[-10pt]
$\sigma_{m}$ & $7.6 \times 10^{-8}$ g/cm$^{2}$ & $D_{ac}$ & $6.8$ eV \\[2pt]
\hline & & & \\[-10pt]
$\hbar \, \omega_{O}$ & $164.6$ meV & $D_{O}$ & $10^{9}$ eV/cm \\[2pt]
\hline & & & \\[-10pt]
$\hbar \, \omega_{K}$ & $124$ meV & $D_{K}$ & $3.5 \times 10^{8}$ eV/cm \\[2pt]
%\hline & & & \\[-10pt]
%
%$\hbar \, \omega_{op-ac}$ & $55$ meV & $D_{f}$ & $5.14 \times 10^{7}$ eV/cm \\[2pt]
\hline
\end{tabular}

\end{table}

\begin{table}[!ht]
\centering
\caption{Physical parameters for the scattering rates related to the substrates.}
\label{Cost_fis_sub}
\begin{tabular}{||c||cc||}
\cline{2-3}
\multicolumn{1}{c||}{} & \multicolumn{1}{c}{SiO$_2$} & \multicolumn{1}{c||}{h-BN} \\
\hline
$\hbar\omega_{op-ac}$ & 55 meV  & 200 meV \\ [2pt]
\hline
$D_f$ & 5.14 $\times 10^7$ eV/cm  & 1.29 $\times 10^9$ eV/cm \\ [2pt]
\hline
$n_i$ & 2.5 $\times 10^{11} $ cm$^{-2}$ & 2.5 $\times 10^{10} $ cm$^{-2}$ \\
\hline
$\kappa_{bottom}$  & 3.9 &  3\\[2pt]
\hline
\end{tabular}
\end{table}
We look for spatially homogeneous solutions to Eq.~\eq{transport} under  a constant applied electric field.
In such a case the transport equation reduces to
\begin{eqnarray}
&&
\dfrac{\partial f(\te,\bk)}{\partial \te} - 
\dfrac{e}{\hbar} \, \mathbf{E} \cdot \nabla_{\bk} f(\te,\bk) =
\int S(\bk', \bk) \, f(\te,\bk') \left( 1 - f(\te,\bk) \right) d \bk' \nonumber
\\
&&
\hspace{106pt} -
\int S(\bk, \bk') \, f(\te,\bk) \left( 1 - f(\te,\bk') \right) d \bk' \p
\label{bulk}
\end{eqnarray}
As initial condition, we take a Fermi-Dirac distribution, 
$$
f(0, \bk) = \dfrac{1}{1 + \exp \left( \dfrac{\eps(\bk) - \varepsilon_F}{k_{B} \, T} \right)} ,
$$
where $T = 300\, \, $ K is the room lattice temperature which will be kept constant.

\section{DSMC method}
We have  also used,  for solving the transport equation \eq{transport}, the ensemble DSMC method recently proposed in \cite{RoMaCo}.
The $\bk$-space is approximated by the set $[-k_{x \, max}, k_{x \, max}] \times [-k_{y \, max}, k_{y \, max}]$ with $k_{x \, max}$ and $k_{y\, max}$ such that the number of electrons  with a wave-vector $\bk$ outside such a set is practically negligible. The  $\bk$-space  is partitioned into a uniform rectangular grid. We shall denote by $C_{ij}$ the generic cell of the grid centered at the 
$\bk_{ij}$ wave-vector. 

The distribution function is approximated with a piecewise constant function in each cell. Initially the $n_P$ particles used for the simulation are distributed in each cell according to the equilibrium Fermi-Dirac distribution. 

The motion of each particle alternates free-flight and scattering. The latter is the most involved and delicate part and in graphene it is particularly important to include the PEP. This implies a heavy computational cost and, more importantly, requires a continuous update of the distribution function. 

In the standard approach the free-flight is performed  according to the semiclassical equation of motion
\begin{eqnarray}
\hbar \dot{\bk} = - e \,  \mathbf{E}.
\end{eqnarray}

The time interval $\Delta \te$ is chosen for each particle in a random way by
\begin{eqnarray}
\Delta \te = - \frac{\ln \xi}{\Gamma_{tot}}, \label{time_free_flight}
\end{eqnarray} 
$\xi$ being a random number with uniform distribution in the interval $[0,1]$ and 
$\Gamma_{tot}$ being the total scattering rate (see for example \cite{JacLug}) 
$$
\Gamma_{tot}  = \Gamma_{ac} (\eps (\te) ) + \Gamma_{op} (\eps (\te) ) + \Gamma_{K} (\eps (\te) ) + \Gamma_{imp} (\eps (\te) ) + \Gamma_{ss} (\eps (\te) ). 
$$
$\Gamma_{ss}$, called {\em self-scattering rate},  is the scattering rate associated to a fictitious scattering that does not change the state of the electron. It is introduced so that    $\Gamma_{tot}$ is constant leading to the simple relation (\ref{time_free_flight}). To fix the value of $\Gamma_{tot}$ one considers the range of the energy involved in the simulation and takes the maximum value $ \Gamma_M$ of the sum 
$\Gamma_{ac} + \Gamma_{op} + \Gamma_{K} +  \Gamma_{imp}$. $\Gamma_{tot}$ is then set equal to $\alpha \Gamma_M$ with $\alpha > 1$ a tuning parameter, e.g. $\alpha = 1.1$. 

Since the range of  $\Gamma_{ac}$,  $\Gamma_{op}$,  $\Gamma_{K}$, $ \Gamma_{imp}$  can be very large, in order to reduce the  computational cost,
a good variant is to use a variable  $\Gamma_{tot}$ which depends on the energy $\eps (\te)$ of the considered  particle at the current  time $\te$
$$
\Gamma_{tot} = \alpha \left(\Gamma_{ac} (\eps (\te) ) + \Gamma_{op} (\eps (\te) ) + \Gamma_{K} (\eps (\te) ) \right).
$$
 We will use this procedure and set $\alpha = 1.1$ in our simulations. 

After the free-flight the scattering is selected randomly according to the values of the transition rates, and PEP  is taken into account as in \cite{Lugli}. Once the state after the scattering has been determined, let us denote  by $\bk'$ its wave-vector, the initial state is changed or left the same with a rejection technique:  a random number $\xi$ is generated uniformly in $[0, 1]$ and if $\xi < 1 - f(\bk')$ the transition is accepted,  otherwise it is rejected. Then,
according to the angular distribution of the scattering rate, a rejection method allows to select the angular dependence of the wave-vector after the scattering event.

At  fixed times the momentum, velocity, energy of each electron are recorded and the mean values are evaluated along with the distribution of electrons among the cells in the $\bk$-space in order to follow the time evolution of the system.  

The maximum number $n_{ij}^{*}$ of simulated particles accommodated in each cell  is easily evaluated (see \cite{Lugli}). Let $N_{ij}$ be the number of real particles in the cell $C_{ij}$ and let $n_{ij}$ be the number of simulated particles in the  same cell.  Let $A$ be the area of the sample and  let $N$ be the number of real particles 
in the sample, $N = \rho A$. By observing that $N/n_p$ is the statistical weight of each particle entering the simulation and taking into account the condition
$0 \le f \le 1$, one has
\begin{eqnarray}
n_{ij} &=& \frac{N_{ij}}{N} n_P = \frac{n_P}{N }\frac{2}{(2 \pi)^2} A \int_{C_{ij}} f \, d\, {\bf k} \le \frac{n_P}{N } \frac{2}{(2 \pi)^2} A \int_{C_{ij}} d\, {\bf k} \nonumber\\
& = &\frac{2}{(2 \pi)^2} \mbox{meas} ({C_{ij}}) \frac{n_P}{N} A = \frac{2}{(2 \pi)^2} \mbox{meas} ({C_{ij}}) \frac{n_P}{\rho} = n_{ij}^{*}, 
\end{eqnarray}
where  $\mbox{meas} ({C_{ij}})$ is the measure of the cell $C_{ij}$. Of course $n_{ij}^{*}$ is not in general an integer, therefore rounding errors are introduced. Usually the problem is solved by using a number of simulated particles $n_P$ great enough to make  such errors negligible. The convergence of the procedure is often checked just by comparing the results with different $n_P$. 

The main concern with the procedure delineated above is that, according to the semiclassical approximation, the compatibility with Pauli's exclusion principle of the positions occupied during the free flight is not checked. {\em It may occur that the particle at the end of the free-flight reaches a cell in the $\bk$-space already fully occupied making the occupation number greater than one} (see \cite{RoMaCo}).

The fact that, for high values of the Fermi energy, the maximum occupation number can greatly exceed the maximum one is of course unphysical, although the average quantities could be acceptable according to the large number law.
Even if the scattering can redistribute  the particles among  the cells, in general it is not  possible to eliminate the presence of anomalous occupation numbers. 
 
For overcoming the problem, in \cite{Reggiani} it has been proposed to apply the rejection technique not only to the scattering event but also at the end of each free-flight. However,  even implementing this  variant, the same drawbacks are still present as shown in \cite{RoMaCo}.

In order to avoid such a difficulty,  in  \cite{RoMaCo} the following approach has been proposed.   The crucial point in the previous procedure is the  step concerning the free-flight. If we go back to the original transport equation, we can use a splitting scheme to avoid unphysical results. The basic idea is to reformulate the splitting method in terms of a particle method. 

In  a time interval $\Delta \te$, first we solve  the drift part of the equation corresponding to the free-flight  in the analogous DSMC approach, 
\begin{equation}
\frac{\partial f (\te,\bx,\bk) }{\partial \te}   - \frac{e}{\hbar} {\bf E} \cdot \nabla_{\bk} f (\te,\bx,\bk)
= 0,
\label{transport_drift}
\end{equation}
taking as initial condition the distribution at time $\te$, 
and then the collision part
\begin{equation}
\frac{\partial f (\te,\bx,\bk) }{\partial \te} = \left. \dfrac{d f}{d \te}(\te,\bx,\bk) \right|_{coll} \sv
\label{transport_coll}
\end{equation}
taking as initial condition the solution of Eq.~\eq{transport_drift}. The global procedure gives a numerical approximation of $f (\te + \Delta \te,\bx,\bk)$ up to first order in $\Delta \te$.
The solution of Eq.~\eq{transport_drift} is just a {\em rigid} translation of the distribution function along the characteristics and can be reformulated from a particle point of view as a free-flight of the same duration for each electron. In this way, the cells in the $\bk$-space are moved of the displacement vector $\hbar \Delta \bk = - e \,  \mathbf{E} \, \Delta \te$ but without changing the occupation number of the cells themselves. To avoid  considering a computational domain as too large,  we adopt a Lagrangian approach and  move the grid by adapting it to the new position of the cells instead of moving the cells. 

Eq.~\eq{transport_coll} is solved by considering a sequence of collision steps for each particle during the time interval $[\te, \te + \Delta \te]$ in a standard way: choice of the scattering, including also the self one, and selection of the final state. Since the collision mechanisms take into account PEP, the occupation number cannot exceed the maximum occupation number  in this second step as well.
Hence, neither the drift nor the collision step give rise to the possibility of having,  in a single cell, more particles than the maximum occupation number.    

The overall scheme  is a hybrid approach which furnishes a first order in time approximation of the distribution function. Average quantities can be evaluated as well by taking the mean values of the quantities of interest, e.g. velocity and energy. 

\section{Numerical results}

We consider a surface impurity density according to Table \ref{Cost_fis_sub}.
 The simulations are performed at several values of the electric field and Fermi energy. 
 
 In order to validate the simulation approach, we numerically solve, in the case of constant $d$, the Boltzmann equation by using also a discountinuos Galerkin (DG) method (see \cite{CoMajRo} for the details) obtaining an excellent agreement. 
 
 $d$ should of the order of few angstroms. In the literature a range from 0 to 1 nm is considered. At variance with \cite{CoMajRo} $d$ is considered a random variable. Therefore in the simulation whenever a  scattering with impurities occurs $d$ is generated according to the chosen distribution. For comparison, the cases with fixed $d$ are also shown along with the solutions obtained by the DG-method.

 Mass conservation implies that the charge density $\rho$, given by 
\begin{eqnarray}
\rho  =  \dfrac{2}{(2 \, \pi)^{2}} \int f(\te, \bk)  \, \: d \bk \sv
\end{eqnarray}
 is constant in time.

We choose a reference frame  such that only the $x-$component of the 
electric field is different from $0$; therefore only the $x-$component of the mean velocity is relevant. 
10$^{5}$  particles have been used for the DSMC. 

\begin{figure}
\begin{center}
\includegraphics[width=0.5\textwidth]{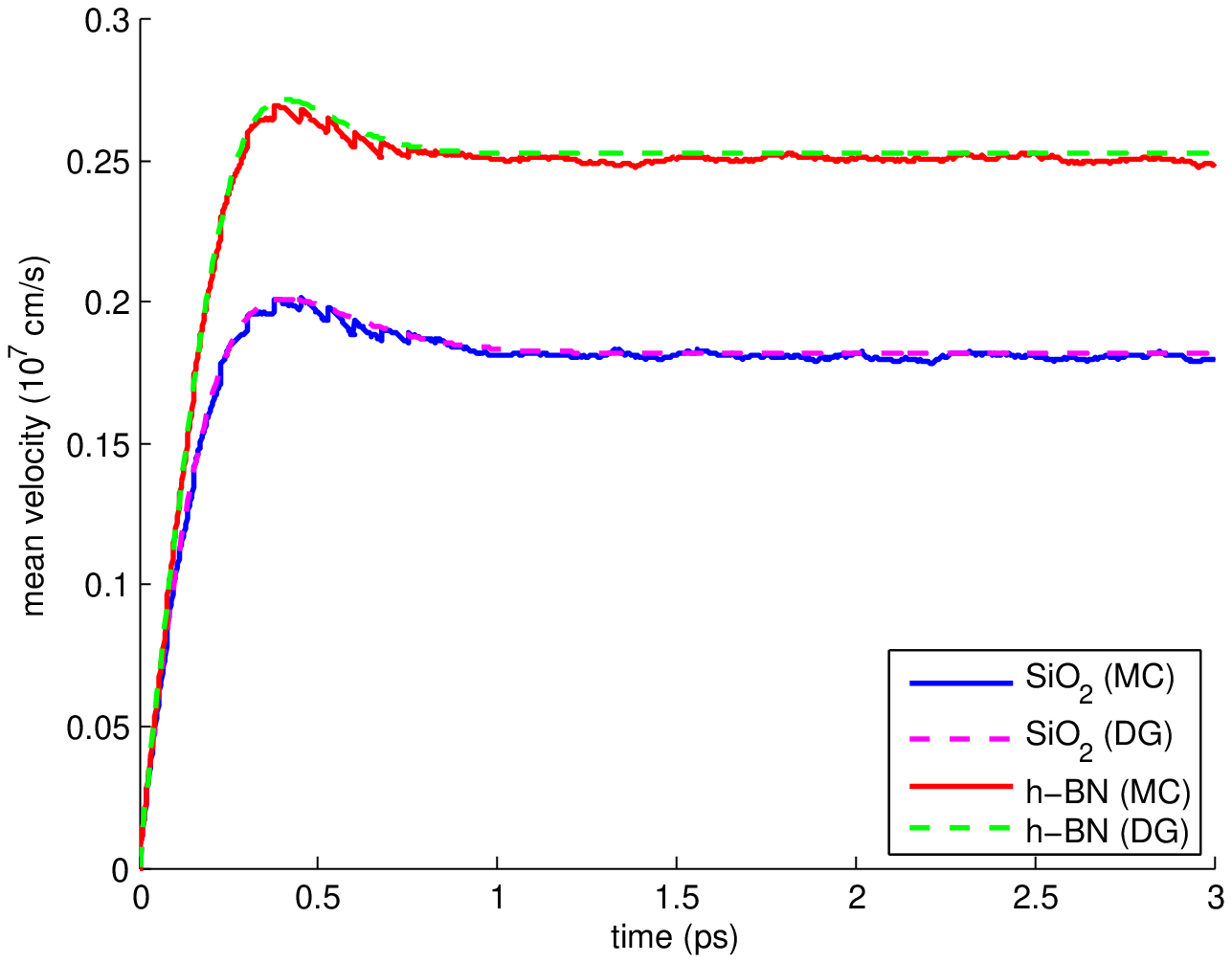}\\[10pt]
\includegraphics[width=0.5\textwidth]{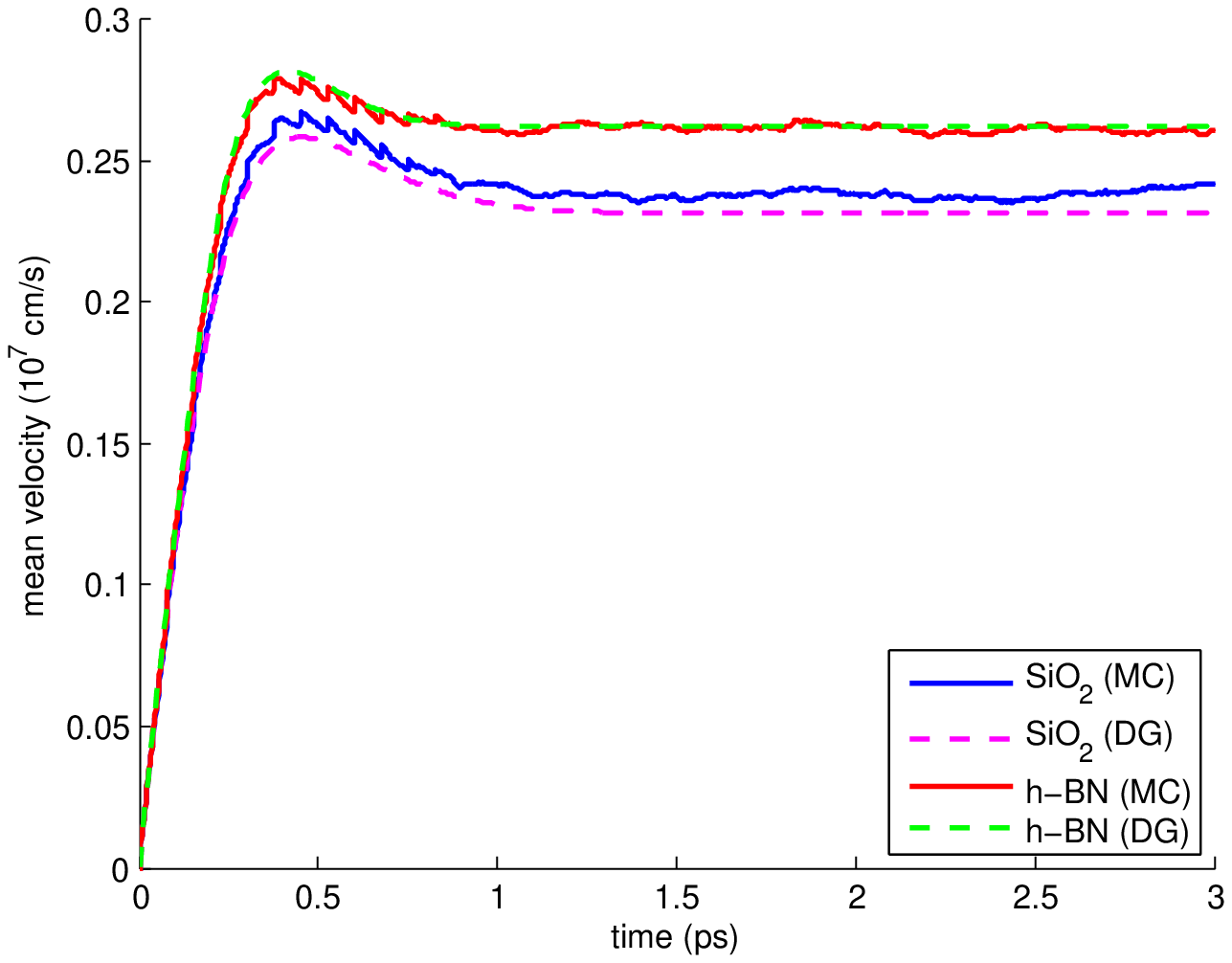}\\[10pt]
\includegraphics[width=0.5\textwidth]{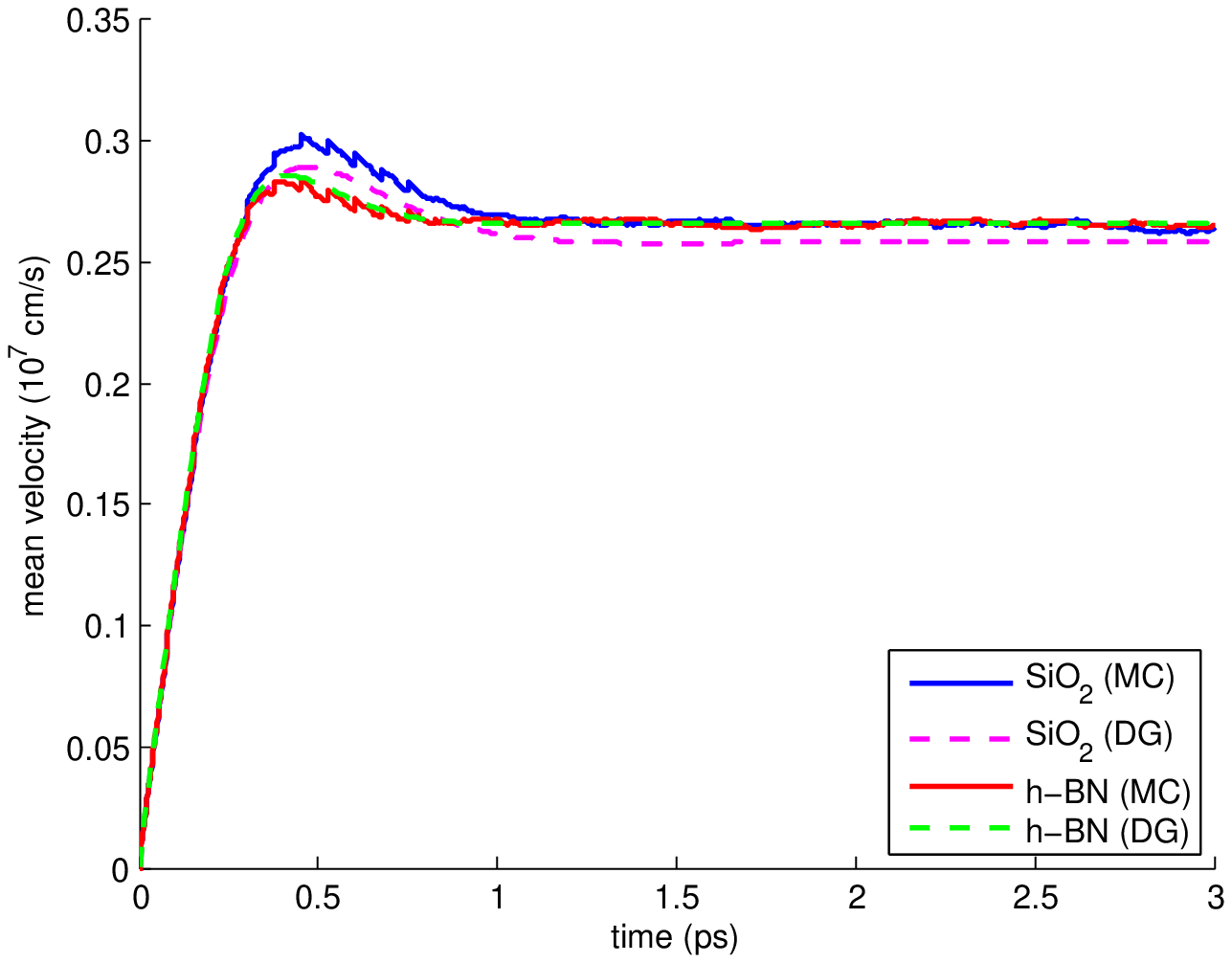} 
\caption{
Comparison of the average velocity  versus time for $d =0$ (top) , $0.5$, $1$ (bottom) nm
 in the case of an applied 
electric field of 5 kV/cm and Fermi energy $\varepsilon_F = 0.4$ eV.
\label{confronto3_5kV}}
\end{center}
\end{figure}

\begin{figure}
\begin{center}
\includegraphics[width=0.5\textwidth]{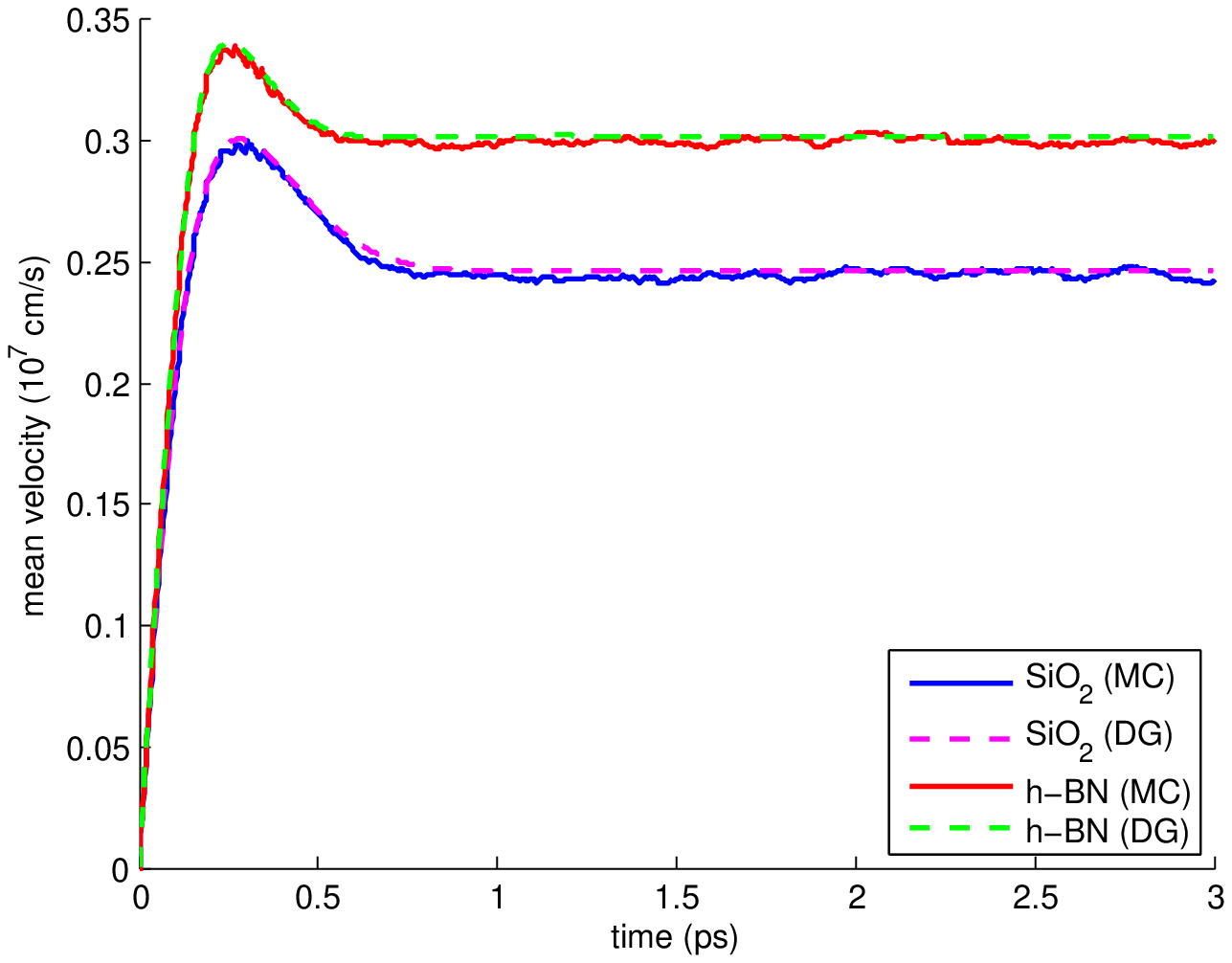}
\includegraphics[width=0.5\textwidth]{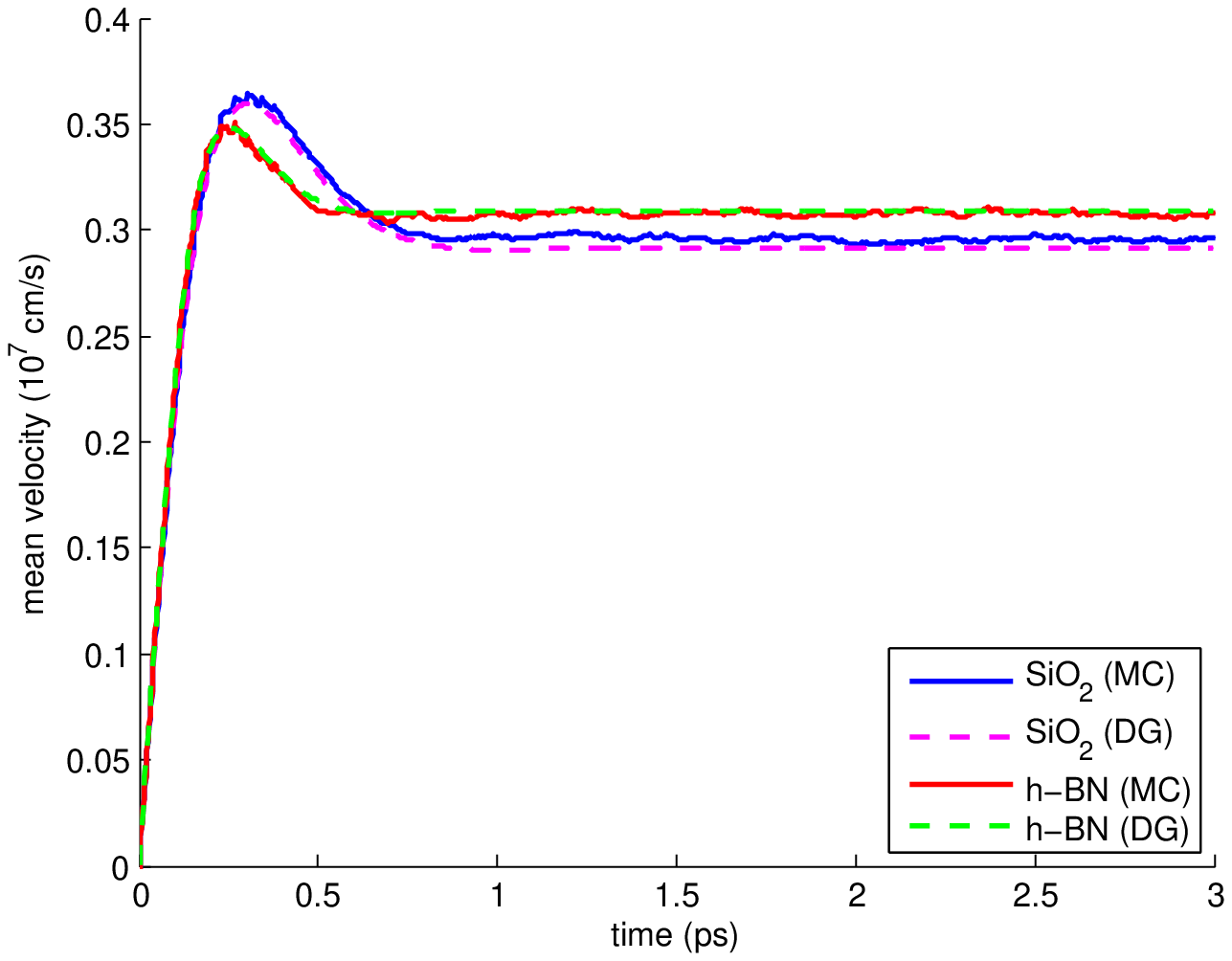}\\[10pt]
\includegraphics[width=0.5\textwidth]{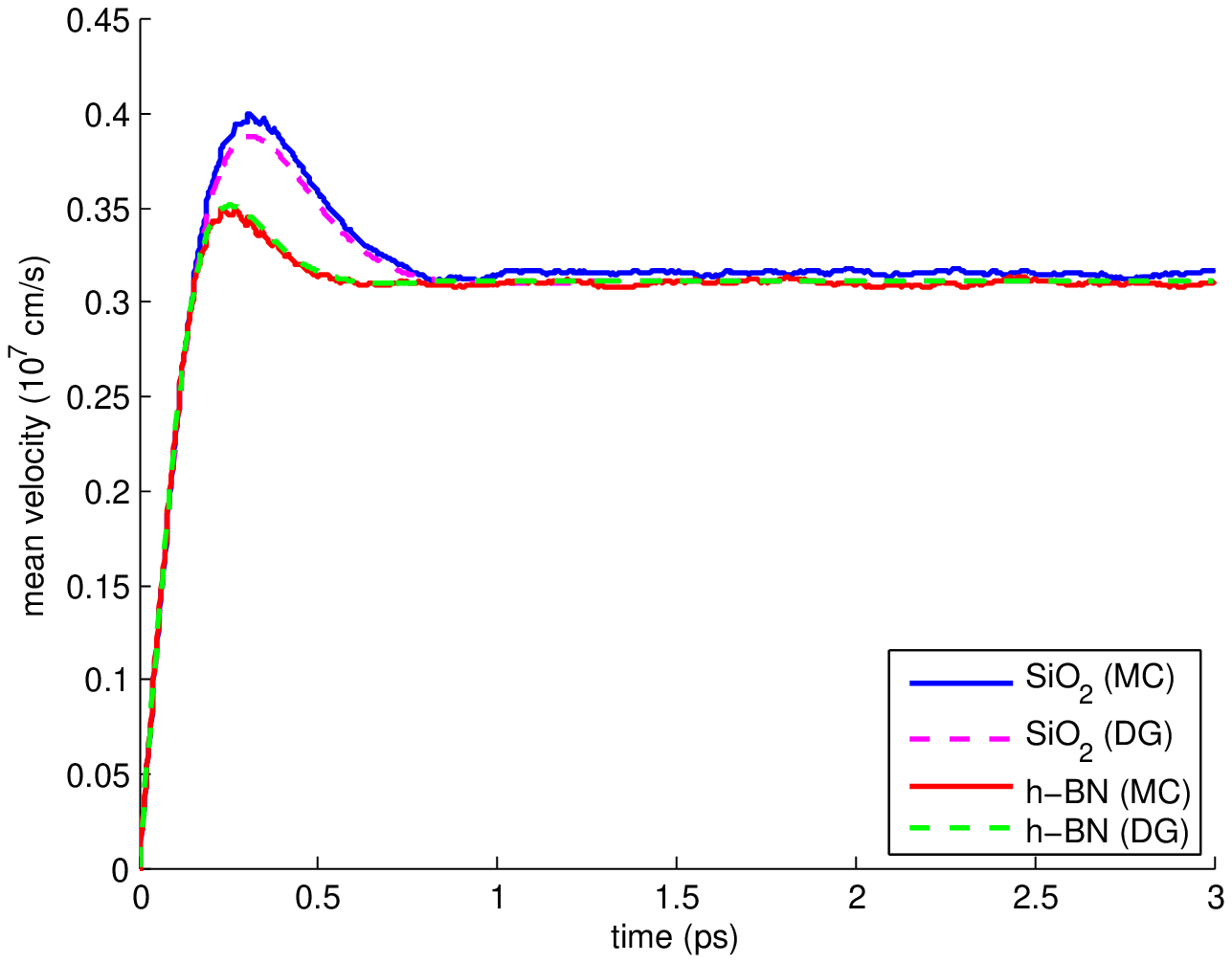} t
\caption{
Comparison of the average velocity  versus time for $d =0$ (top) , $0.5$, $1$ (bottom) nm
 in the case of an applied 
electric field of 10 kV/cm and Fermi energy $\varepsilon_F = 0.4$ eV. 
Both the results obtained by using the Monte Carlo (MC) and the discontinuous Galerkin methods (DG)  are reported. 
\label{confronto3_10kV}}
\end{center}
\end{figure}

\begin{figure}
\begin{center}
\includegraphics[width=0.5\textwidth]{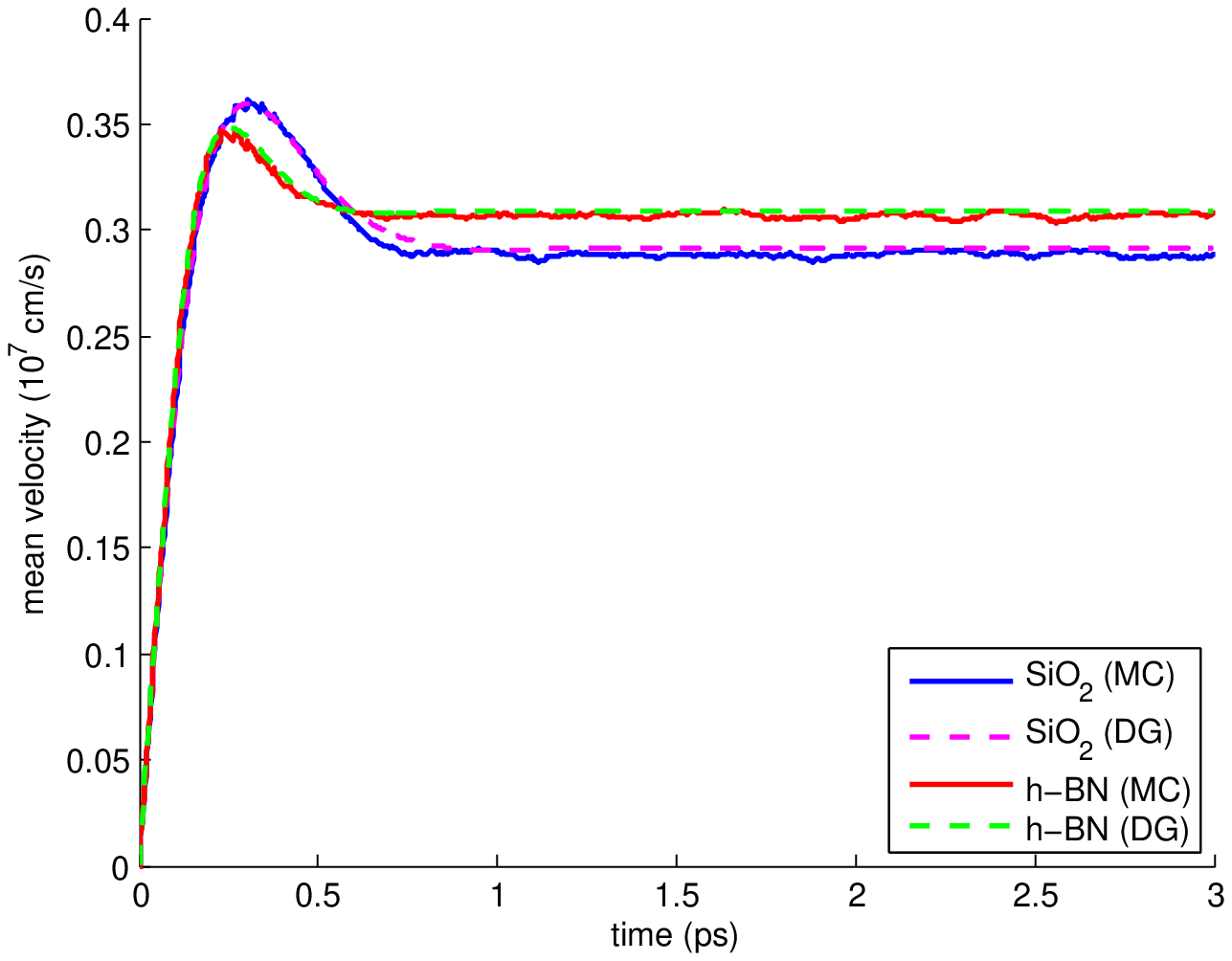}\includegraphics[width=0.5\textwidth]{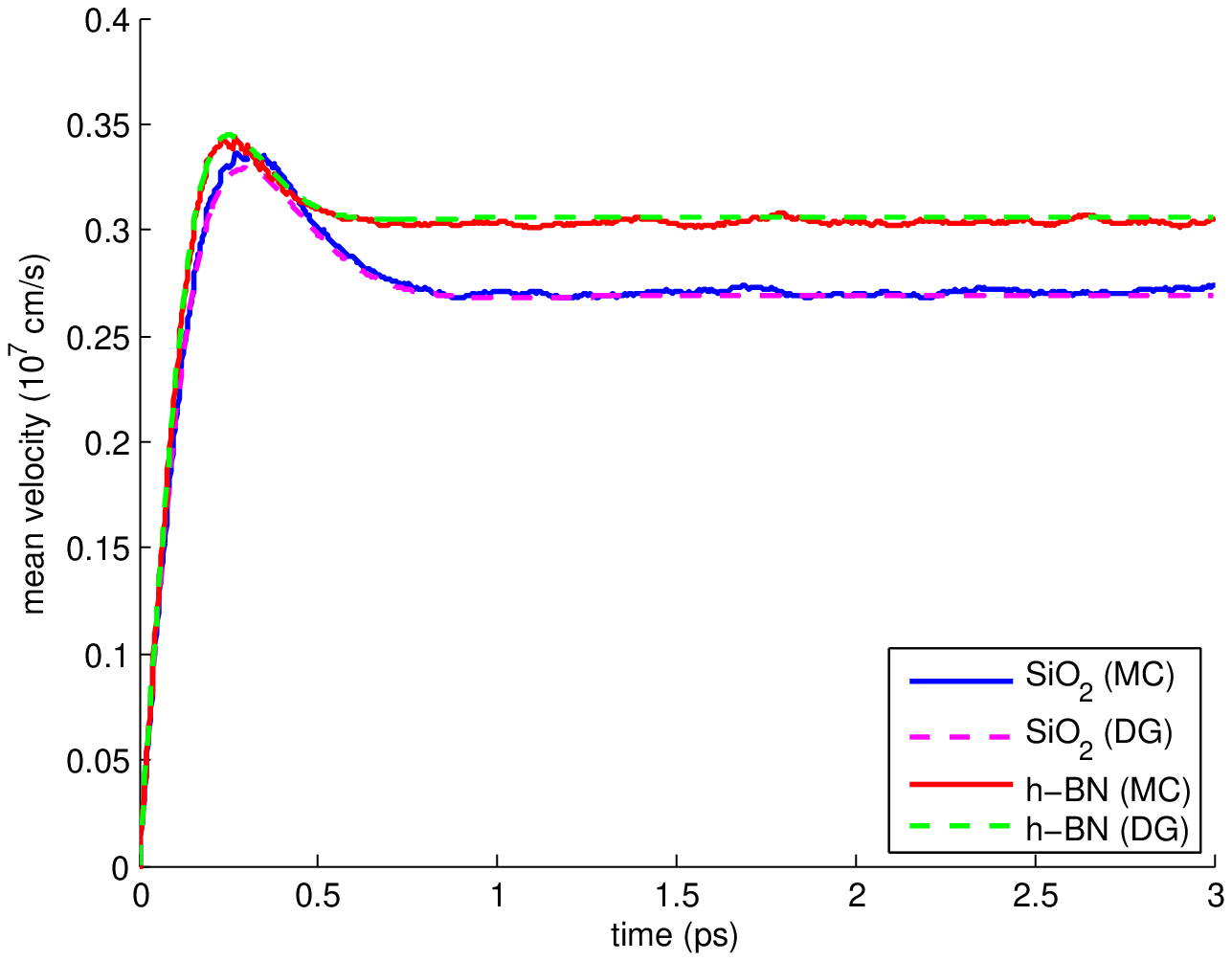} \\[10pt]
\includegraphics[width=0.5\textwidth]{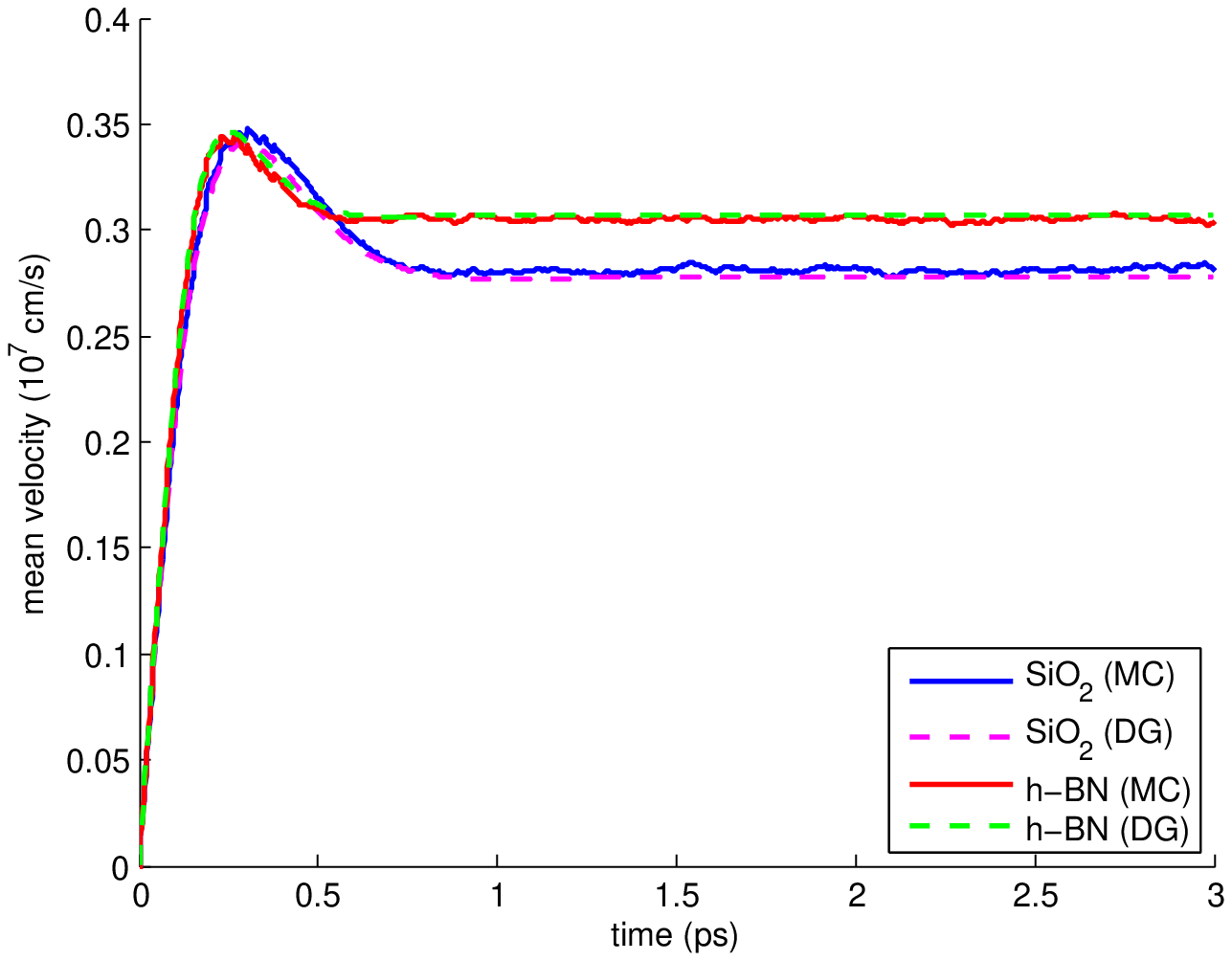}\includegraphics[width=0.5\textwidth]{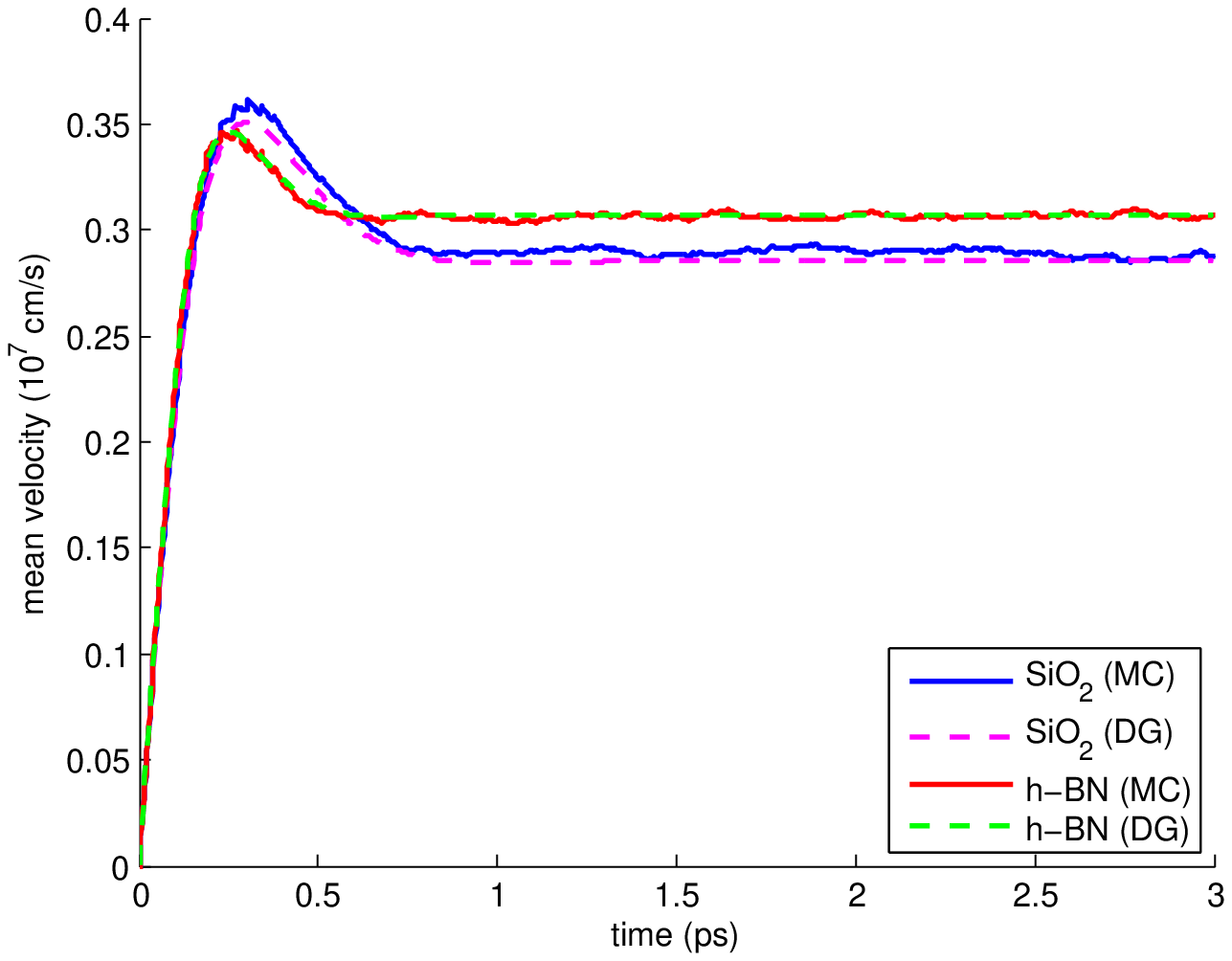}
\caption{
Comparison of the average velocity versus time in the case of an applied 
electric field of 10 kV/cm and Fermi energy $\varepsilon_F = 0.4$ eV by considering different distribution for $d$:
uniform (top left), $\Gamma(2,0.5)$  (top right),  $\Gamma(3,0.5)$ (bottom left),
$\Gamma(4,0.5)$ (bottom right). In the results obtained with the discontinuous Galerkin methods (DG)  we have assumed $d$ equal to the mean values of the corresponding distribution rescaled by the factor 0.2 nm.
\label{confronto2_dvar_1}}
\end{center}
\end{figure}

In Figs.~\ref{confronto3_5kV}-\ref{confronto2_dvar_1}, we show the numerical results of the average velocity 
$\mathbf{v}$ defined as 
\begin{equation}
\mathbf{v}(\te) = \dfrac{2}{(2 \, \pi)^{2} \, \rho} \int f(\te, \bk) \dfrac{1}{\hbar} \, \nabla_{\bk} 
\, \eps(\bk) \: d \bk \sv
%\quad
%W (\te) = \dfrac{2}{(2 \, \pi)^{2} \, \rho} \int f(\te, \bk) \, \eps(\bk) \: d \bk \p
\end{equation}
The velocity is related to the current $\mathbf{J}$ by the relation
$$
\mathbf{J} = - e  \rho \mathbf{v} 
$$
and, in turn, $\mathbf{v}$ is related to the mobility $\mu (\mathbf{E})$ as follows
$$
\mathbf{v} = \mu (\mathbf{E} ) \mathbf{E}. 
$$
Therefore, from the analysis of the average velocity it is possible to estimate the effect of the impurities on the mobility.
It is expected that the scattering with the remote impurities leads to a degradation of the mobility depending on the specific material. 

First we have assessed the general performance of the different materials, by a comparison of the average velocity for three different values of $d$ kept constant. 

We can observe that the values of the average velocity and energy become  lower by reducing the distance  $d$ from the impurities in the oxide,  confirming the degradation of the mobility due to the substrate as  a direct consequence of the additional scatterings with the remote  impurities. For the higher value of $d$,  which is very close to the pristine case, both SiO$_2$ and h-BN produce of course the same effect with a comparable electron velocity. For the intermediate value of $d$ 
h-BN performs better than  SiO$_2$ and this behaviour is even more evident for $d=1$. Therefore,  h-BN gives a better high-field mobility, in  qualitative agreement with the low field analysis in \cite{Hirai}.

\begin{figure}
\begin{center}
\includegraphics[width=0.6\textwidth]{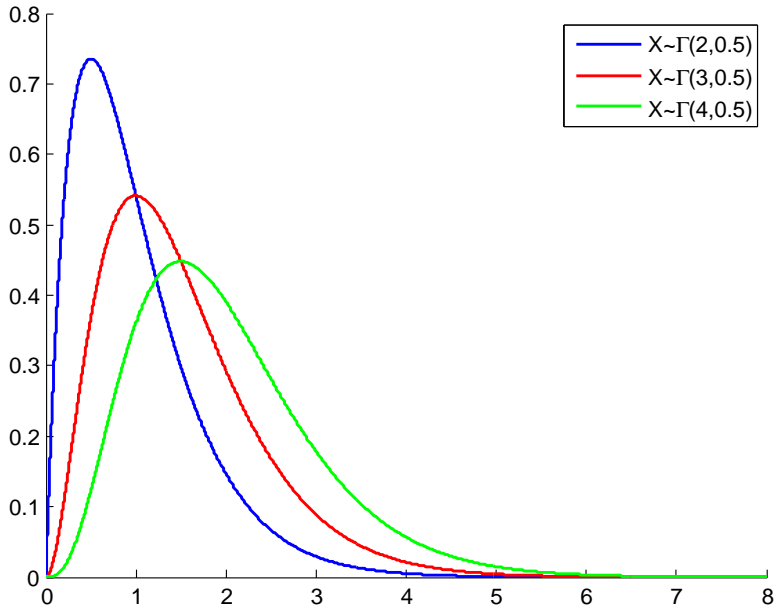} 
\caption{ Plot of the $\Gamma (\alpha, \lambda)$ distribution with  $\lambda = 0.5$ and $\alpha = 2, 3, 4$. Note that the probability to generate a number greater than 5 is practically
zero. 
\label{chi2}}
\end{center}
\end{figure}

The previous results, however, do not take into account the intrinsic noise in the location of the impurities. In order to assess its effect on the high-field mobility, we have performed some simulations with a random $d$ generated, in each scattering involving impurities, according to a prescribed probability distribution (see Fig. \ref{confronto2_dvar_1}). First we have considered a uniform distribution in the interval $[0, 1]$ (in nm). The results are similar to the case with constant  $d = 0.5$ nm and this can be explained by observing that 0.5 is the expectation value. Then we have considered a $\Gamma (\alpha, \lambda)$ distribution 
$$
f(x) = \left\{
\begin{array}{ll}
\dfrac{1}{\lambda \Gamma (\alpha)} x^{\alpha - 1} e^{x/ \lambda} & x > 0\\
0 & x \le 0
\end{array}
\right.
$$
where $\Gamma (\alpha)$ is the Euler gamma function. We have used the values $\lambda = 0.5$ and $\alpha = 2, 3, 4$  (see Fig. \ref{chi2})  and rescaled $d$ by a factor 0.5 nm in order to have a values less than 1 with very high probability, as confirmed by the simulation. 

In order to validate our findings, the results obtained by using the DG-method in \cite{CoMajRo} but with a valued of $d$ set equal to the mean values of the considered distribution ($\alpha \lambda$ for the $\Gamma (\alpha, \lambda)$ one) rescales by the factor 0.2 nm. The agreement is still excellent.  

We would like to conclude by observing that both the materials 
 seem only slightly  influenced by the stochastic effect related to the randomness of the impurity positions. 

\section{Conclusions}
An analysis of the high-field mobility has been performed in graphene on a substrate by a new DSMC approach which properly takes into account the Pauli exclusion principle. The same substrates as in \cite{Hirai} have been considered but with the  more elaborate model for the charge-impurities scattering proposed in  \cite{Hwang2007b}.  Moreover, also the random distribution of the depth of the impurities implanted in the oxide has been taken into account 
and described with several theoretical probability distributions.

The differences among the average velocities for the considered substrates  are in agreement with the expected effects and confirm a degradation of the mobility.  As already found out in \cite{Hirai} for the low field mobility, h-BN reveals a better substrate than SiO$_2$ because produce a smaller degradation also in the high-field mobility.

% Non-BibTeX users please use

\end{document}